\documentstyle[12pt]{article} 
\begin{document}
~\hspace*{11.5cm} ADP-AT-99-2\\
 ~\hspace*{10.9cm} M.N.R.A.S., submitted\\[2cm]

\begin{center}
{\large \bf The physical parameters of Markarian 501\\ during flaring 
activity}\\[1cm]

W. Bednarek$^{1,2}$ and R.J. Protheroe$^1$ \\
$^1$Department of Physics and Mathematical Physics\\
The University of Adelaide, Adelaide, Australia 5005.\\
rprother@physics.adelaide.edu.au\\
$^2$University of \L\'od\'z, 90-236 \L\'od\'z, 
ul. Pomorska 149/153, Poland.\\
bednar@krysia.uni.lodz.pl
\end{center} 

\begin{abstract}
We determine the physical parameters (magnetic field and Doppler
factor) of the homogeneous synchrotron self-Compton model allowed
by the observed X--ray to gamma-ray spectra and variability of
Markarian~501 during the 15--16 April 1997 flaring activity.  We
find that magnetic fields between 0.07~G and 0.6~G and Doppler
factors between 12 and 36 could fit (depending on observed
variability time scale) these observations.  We take account of
photon-photon pair production interactions of gamma-ray photons
occurring both inside the emission region and during propagation
to Earth and find these to be extremely important in correctly
determining the allowed model parameters.  Previous estimates of
the allowed parameter space have neglected this effect.  Future
multi-wavelength campaigns during strong flaring activity,
including observations from optical to TeV gamma-rays, should
enable the physical parameters to be further constrained.\\

\noindent {\bf keywords:}
galaxies: active -- quasars: jets -- radiation mechanisms: gamma
rays -- galaxies: individual: Mrk~501\\
\end{abstract}

\section{Introduction}
Four BL Lac objects have been detected in the TeV energy range:
Mrk~421 (Punch et al.\ 1992), Mrk~501 (Quinn et al. 1996),
1E~S2344+514 (Catanese et al.\ 1998), and PKS~2155-304 (Chadwick
et al.\ 1998).  In two of these, Mrk 421 and Mrk 501, the high
level of $\gamma$-ray emission enabled the spectrum to be
measured up to $\sim 10$ TeV (Zweerink et al.\  1997, Aharonian et
al. 1997, Samuelson et al.\ 1998, Djannati-Atai et al.\  1998,
Hayashida et al.\ 1998).  In the case of Mrk 421, the spectrum can
be adequately described by a single power law between 0.3 - 10
TeV.  However the spectrum of Mrk 501 shows clear curvature over
a similar energy range (Krennrich 1998).  Recently, the spectrum
of Mrk 501 has been measured up to 24 TeV by the HEGRA telescopes
(Konopelko et al.\ 1999, Krawczynski et al.\  1999).

The $\gamma$-ray emission of these two BL Lacs shows very rapid
variability.  For Mrk 421, variability on a time scale
as short as $\sim 15$ minutes has been reported (Gaidos et
al. 1996).  In the case of Mrk 501, variability on a time scale
of a few hours was observed during the 1997 high level of
activity (Aharonian et al.\ 1998, Quinn et al.\ 1999 cited by 
Krennrich et al.\ 1998), and there is
some evidence of variability on a time scale of 20 minutes
(Aharonian et al.\ 1998).  

The TeV $\gamma$-ray flares are simultaneous with the X-ray
flares. In the case of Mrk 501, during the 16 April 1997 flare
the X-ray spectrum was observed by the Beppo-SAX observatory up
to $\sim 200$ keV (Pian et al.\ 1998), and during the same high
state OSSE observations made between 9 to 15 April 1997 (Catanese
et al. 1997) show that the energy flux per log energy interval
continues up to $\sim 500$ keV at roughly the same level.

Gamma-ray emission from active galactic nuclei (AGN) is often
interpreted in terms of the homogeneous ``synchrotron
self-Compton model'' (SSC) in which the low energy emission (from
radio to X-rays) is synchrotron radiation produced by electrons
which also up-scatter these low energy photons into high energy
$\gamma$-rays by inverse Compton scattering (ICS) (Macomb et
al.~1995, Inoue \& Takahara~1996, Bloom \& Marscher~1996,
Mastichiadis \& Kirk~1997). In this model all the radiation comes
from this same region in the jet. Such a picture can naturally
explain synchronized variability at different photon
energies.  

The inclusion of photon-photon pair production interactions of
$\gamma$-rays with low energy radiation within the emission
region has been used previously when constraining the physical
parameters of blazars (e.g. Mattox et al.~1993, Dondi \&
Ghisellini~1995).  This was also included in our previous paper
(Bednarek and Protheroe 1997) where we constrained the allowed
parameter space of the homogeneous SSC model, i.e. the magnetic
field in the emission region and the Doppler factor, based on
observations of the 1994 flaring activity in Mrk 421.  More
recently, a similar analysis has been performed for Mrk 501 by
various authors (Kataoka et al.\ 1999, Tavecchio, Maraschi and
Ghisellini 1998).  However, in this recent work absorption on
both the infrared background (IRB) and the internal radiation of
the emission region has been neglected.  In the present paper we
show that inclusion of both these effects is vital in order to
properly determine the allowed parameter space.

In Section 2 we discuss the effect of photon-photon pair
production interactions of $\gamma$-rays with synchrotron radiation
produced inside the emission region, and with IRB
photons during propagation to Earth.  In Section~3 we obtain the
values of $B$ and $D$ allowed by the observed ratio of $\gamma$-ray
to X--ray power, and in Section~4 we discuss additional
constraints arising from the requirement that the radiative
cooling time must be consistent with the observed variability
time scale.  In Section~5, we further narrow down the allowed
model parameters by comparing the predicted TeV $\gamma$-ray
spectrum with that observed, and finally we discuss the implications
of our results.

\section{Absorption of Gamma-Rays by Photon Photon Pair Production }

For propagation of $\gamma$-rays through isotropic radiation the
reciprocal of the mean interaction length for photon--photon pair
production is given by
\begin{equation}
x_{\gamma \gamma}(E_\gamma)^{-1} = {1 \over 8 {E_\gamma}^2}
\int_{\varepsilon_{\rm min}}^{\infty} \, d\varepsilon
\frac{n(\varepsilon)} {\varepsilon^2} \int_{s_{\rm min}}^{s_{\rm
max}(\varepsilon,E_\gamma)} ds \, s \sigma(s),
\label{eq:mpl}
\end{equation}
where $n(\varepsilon)$ is the differential photon number density
and $\sigma(s)$ is the total cross section for photon-photon pair
production (Jauch and Rohrlich 1955) for a centre of momentum
frame energy squared given by $s=2 \varepsilon E_\gamma(1 - \cos
\theta)$ where $\theta$ is the angle between the directions of
the energetic photon and soft photon, and $s_{\rm min} = (2 m_e
c^2)^2$, $\varepsilon_{\rm min} = (2 m_e c^2)^2 /
4E_\gamma$, and $s_{\rm max}(\varepsilon,E_\gamma) =
4\varepsilon E_\gamma$.

We shall now apply these formulae directly to obtain the mean
interaction length in the IRB and, in a modified form, to obtain
the optical depth in the synchrotron radiation produced in the
emission region.

\subsection{Absorption of gamma-rays in the infrared background}

The spectra of $\gamma$-rays from extragalactic objects are modified
by photon photon pair production interactions with the infrared
background, and observations of the $\gamma$-ray spectra of Mrk~421
and Mrk~501 have been used to constrain the intensity of the
infrared background (Stanev and Franceschini 1998, Biller et
al. 1998).  Recently, Malkan and Stecker (1998) have modelled the
infrared background by summing the contributions from different
types of galaxy taking account of evolution.  We use their upper
and lower models which are consistent with recent COBE DIRBE data
(Hauser et al.\ 1998) and COBE FIRAS data (Fixsen et al.\ 1998)
which we use to extend the model of Malkan and Stecker down below
3~$\mu$m.

\begin{figure}
\vspace{13.0cm}
\includegraphics{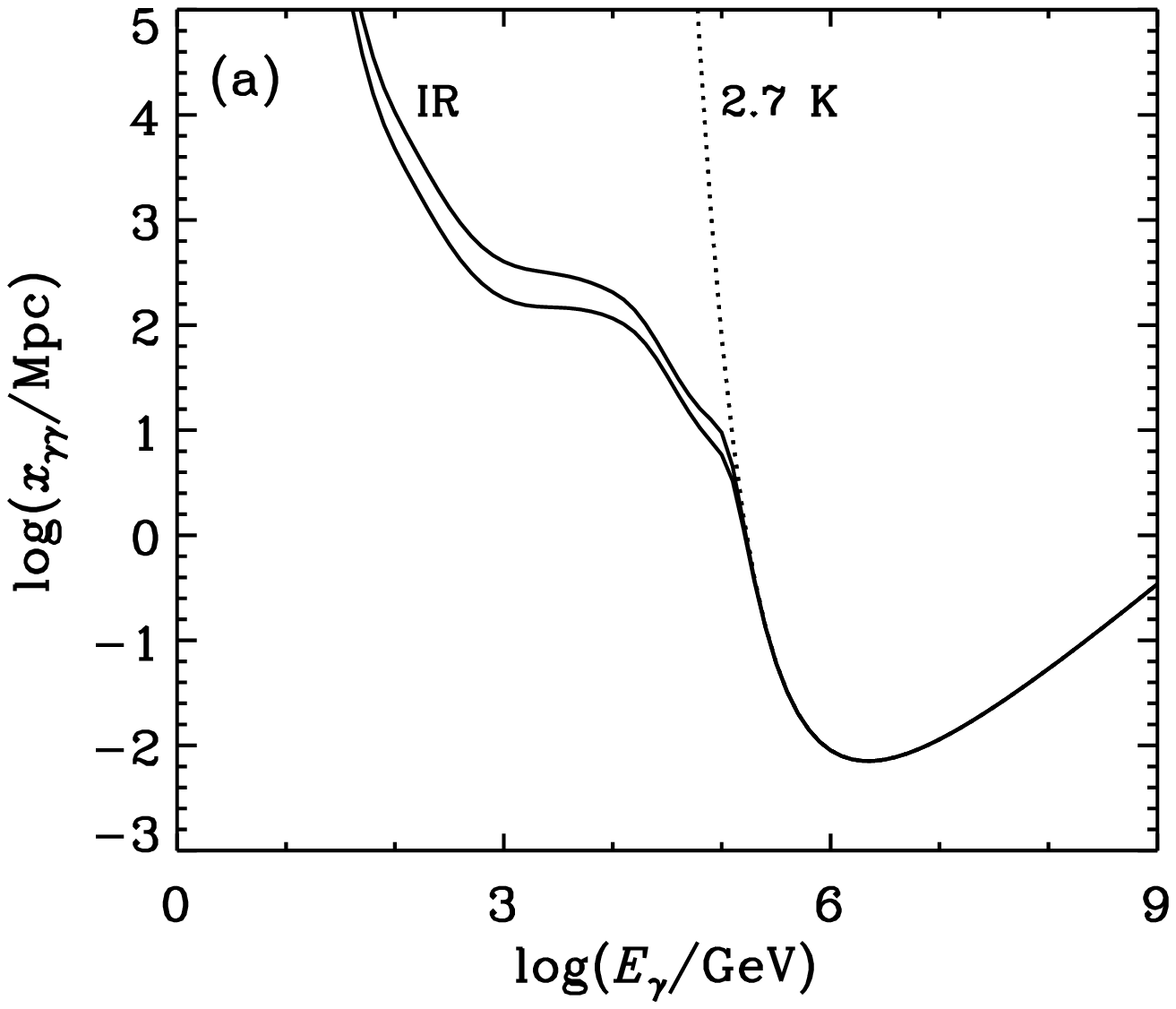}
\includegraphics{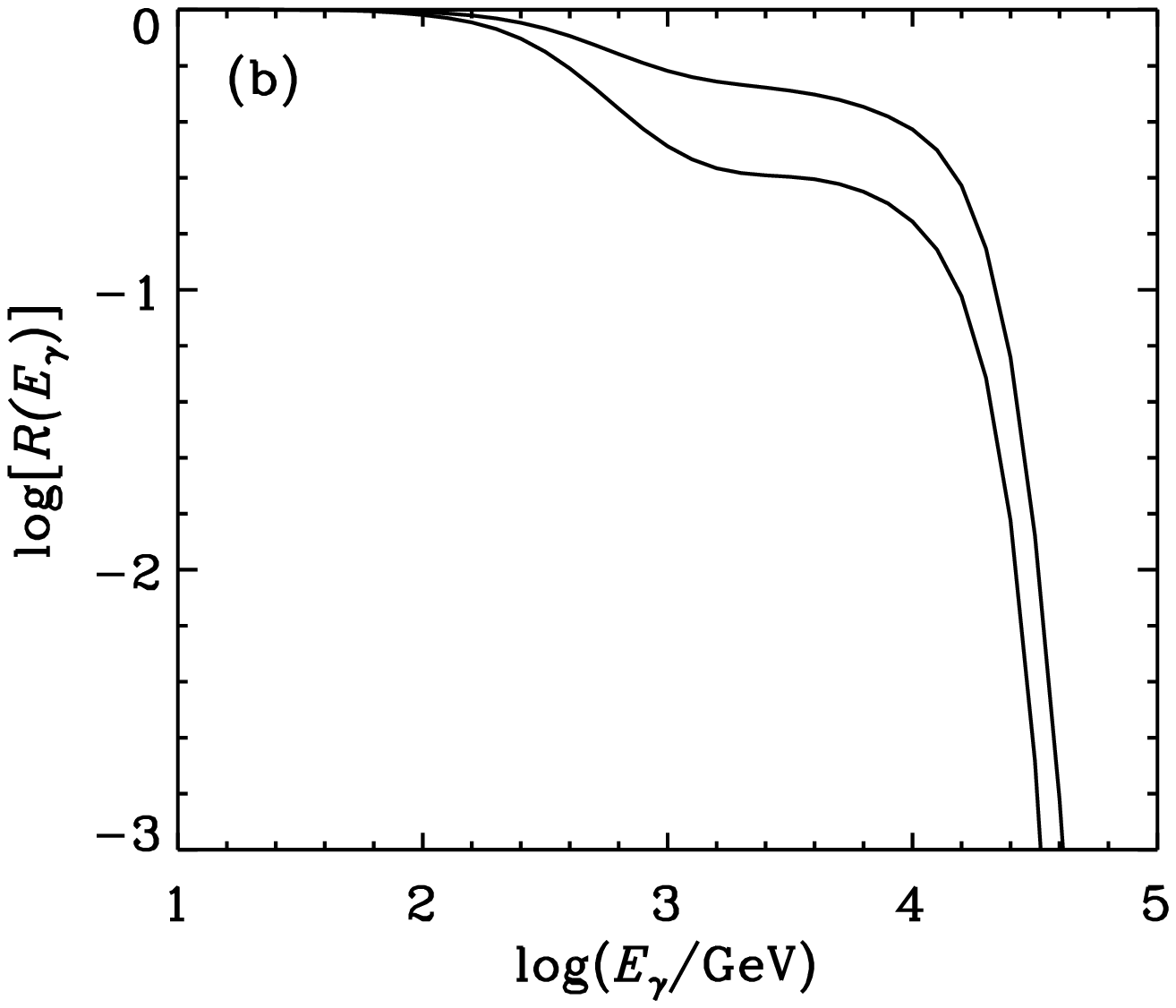}
\caption{(a) Mean interaction length for photon-photon pair
production in the cosmic microwave background plus infrared
background upper and lower models of Malkan and Stecker (solid
curves).  Dotted curve labelled 2.7~K shows mean interaction
length in the cosmic microwave background.  (b) Reduction factor,
$R(E_\gamma)=\exp(-\tau)$ for a source distance of 202 Mpc.}
\label{fig:ir_absn}
\end{figure}

\begin{figure}
\vspace{12.0cm}
\includegraphics{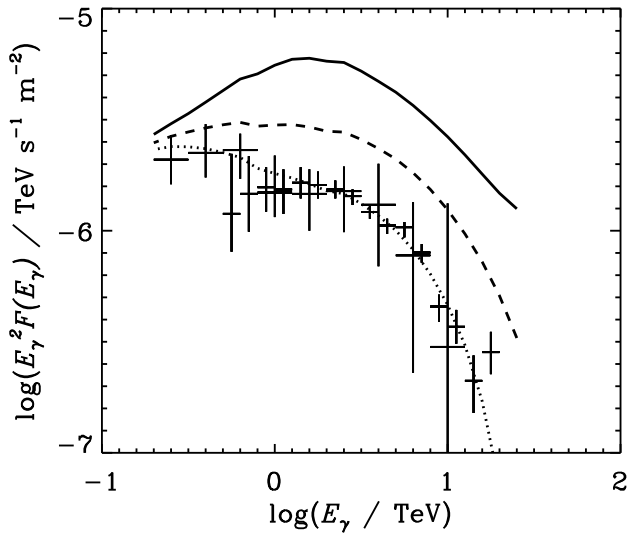}
\caption{Mrk~501 1997 April 15--16 HEGRA data (data binned in
1/10 decade intervals) and CAT data (data binned in 1/5 decade
intervals).  The error bars on the HEGRA data include systematic
errors as given in Fig.~1b of Krawczynski et al.~(1999).  The
dotted curve shows the SED used in this paper, and the solid and
dashed curves show the SED after correction for absorption using
the lower and upper reduction factors of Fig.~1(b).}
\label{fig:tev501high}
\end{figure}

We obtain the mean interaction length, $x_{\gamma \gamma} (E_\gamma)$, 
in the IRB plus cosmic
microwave background radiation (CMBR) and these are shown in
Fig.~\ref{fig:ir_absn}(a).  We also show separately the mean free
path in the CMBR alone.  Our results are in good agreement with
those of Stecker and De Jager (1998) calculated for the same
infrared background models (see Stecker and De Jager for
references to earlier work).  In Fig.~\ref{fig:ir_absn}(b) we
show the reduction factor, $R(E_\gamma)=\exp[-\tau_{{\rm
IR}}(E_\gamma)] \equiv \exp[-d/x_{\gamma \gamma} (E_\gamma)$ for a
source distance of $d=202$~Mpc obtained for Mrk~501 assuming $H_0
= 50$ km~s$^{-1}$~Mpc$^{-1}$ and a redshift of $z=0.033$.
Clearly, absorption in the IR background becomes very important
above 400~GeV.  Figure \ref{fig:tev501high} shows the 1997 April
15--16 HEGRA and CAT data together with the approximation used
later in this paper for the high energy part of the spectral
energy distribution (SED).  The figure also shows this SED after
correction for absorption in the infrared using the two reduction
factors of Fig.~\ref{fig:ir_absn}(b).

\subsection{Absorption of gamma-rays in the blob radiation}

Absorption of $\gamma$-rays will also take place on the synchrotron
photons produced inside the emission region.  For the spectrum of
these target photons we use a fit to the Beppo-SAX observations
made during the April 15/16 flaring activity (Pian et al.\ 1998),
together with an indication from OSSE Observations made during
the high state in April 1997 (Catanese et al.\ 1997) that the
spectrum continued to $\varepsilon_{\rm s,max} = 500$
keV with approximately the same energy flux per log energy
interval.  The differential photon flux in the optical to X-ray
region observed from Mrk 501 during the 16 April 1997 flare
(photons cm$^{-2}$ s$^{-1}$ GeV$^{-1}$) is approximated by
\begin{eqnarray}
F(\varepsilon) \approx \cases {3\times 10^{-4}\varepsilon^{-1.4} & $
\varepsilon \le 
\varepsilon_{s,1}=2.14\times 10^{-6} {\rm GeV}$,\cr
2.5\times 10^{-5} \varepsilon^{-1.59} & $\varepsilon_{s,1} < \varepsilon \le 
\varepsilon_{s,2}=1.95\times 10^{-5} {\rm GeV}$,\cr
1.86\times 10^{-6}\varepsilon^{-1.83} & $\varepsilon_{s,2} < \varepsilon \le 
\varepsilon_{s,3}=2\times 10^{-4} {\rm GeV}$,\cr
4.37\times 10^{-7}\varepsilon^{-2} & $\varepsilon_{s,3} < \varepsilon \le 
\varepsilon_{\rm s,max}$.\cr}
\label{eq:SED}
\end{eqnarray}

The photon density in the emission region depends on the dimensions
of the source and the assumed Doppler factor.  We assume that
relativistic electrons are confined inside a ``blob''
which moves along the jet with the Doppler factor $D$ and has
magnetic field $B$.  In the homogeneous SSC model the radii of
the emission regions of low energy photons ($r_l$), X-ray photons
($r_X$), and TeV $\gamma$-rays ($r_{\gamma}$) are the
same. This region is constrained by the variability time scale
observed, e.g. in TeV $\gamma$-rays, $t_{\rm var}$,
\begin{eqnarray}
r_l = r_{\gamma} = r_X \approx 0.5 \xi c D t_{\rm var}
\label{eq:radius}
\end{eqnarray} 
where $\xi \le 1$, $\xi = 1$ being the usual assumption.
The differential photon density in the blob frame of 
synchrotron photons is then given by
\begin{eqnarray}
n(\varepsilon') \approx {{4d^2 F(\varepsilon)}\over{c^3 t_{\rm
var}^2 D^4\xi^2}},
\label{eq:photden}
\end{eqnarray}
where $\varepsilon =
D\varepsilon'$ and $\varepsilon'$ are the photon energies in the
observer's and the blob rest frames, and $c$ is the velocity of
light.  

The optical depth $\tau_{{\rm syn}}(E'_\gamma)$ 
in the blob frame for $\gamma$-ray photons with blob-frame
energy $E'_\gamma=E_\gamma/D$ for $e^\pm$ pair production inside
the blob is given by
\begin{equation}
\tau_{{\rm syn}}(E'_\gamma) = {r_\gamma \over 8 {E'_\gamma}^2}
\int_{\varepsilon'_{\rm min}}^{\infty} \, d\varepsilon'
\frac{n(\varepsilon')} {\varepsilon'^2} \int_{s_{\rm min}}^{s_{\rm
max}(\varepsilon',E'_\gamma)} ds \, s \sigma(s).
\label{eq:tau_blob}
\end{equation}
We compute the optical depth by numerical integration of
Eq.~\ref{eq:tau_blob}, and our results can be approximated very
well in the energy range of interest, i.e. 0.1~TeV $< E_\gamma <
100)$~TeV, by
\begin{equation}
\tau_{{\rm syn}}(E_\gamma) \approx 3\times 10^8 D^{-4.8}
E_\gamma^{0.4} t_{\rm var}^{-1}\xi^{-1},
\label{eq:tau_blob_approx}
\end{equation}
for $1 \le D \le 100$ and $10^2$~s~$\le t_{\rm var} \le 10^4$~s.
We note that the $E_\gamma^{0.4}$ energy dependence we obtain is
consistent with equation~(B7) of Svensson~(1987) for a photon
spectral index of 1.6.  From Eq.~\ref{eq:tau_blob_approx}, we
find that $D \gg 13$ to avoid absorption at 1~TeV if the
variability time is 20 minutes and applying $\xi = 1$.  Clearly
absorption by photon-photon interactions inside the blob can not
be neglected.


\section{Constraint from Ratio of Gamma-Ray to X--ray Power}

The spectrum of Mrk 501 shows two clear bumps which, during the
outburst stage, extend up to at least $\sim 500$ keV
(Pian et al.\ 1998, Catanese et al.\ 1997), and up to at least
$\sim 10$ TeV (Aharonian et al.\ 1997, Krennrich et
al. 1999, Djannati-Atai et al.\ 1998, Hayashida et al.\ 1998, see
also Protheroe et al.\ 1998 for review), the highest energy of
$\gamma$-rays observed being 24 TeV (Konopelko et al.\ 1999,
Krawczynski et al.\ 1999).  These multiwavelength observations of
Mrk 501 allow us to define the ratio $\eta$ of the energy flux per 
log energy interval observed
at a chosen $\gamma$-ray energy, $E_\gamma=1$~TeV, at which the
emission is assumed to be due to Compton scattering, to that at 
a chosen X--ray energy, $\varepsilon=2$~keV, at which the
emission is assumed to be due to synchrotron radiation,
\begin{eqnarray}
\eta = ({{dN}\over{dE_\gamma dt}} E_\gamma^2 e^{-\tau_{\rm
tot}(E_\gamma)})/({{dN} \over{d\varepsilon dt}} \varepsilon^2) =
({{dN}\over{dE'_\gamma dt'}} {E'}_{\gamma}^{2} e^{-\tau_{\rm
tot}(E'_\gamma)})/{{dN}\over{d\varepsilon' dt'}}{\varepsilon'}^{2}),
\label{eq:eta}
\end{eqnarray}
where $\tau_{\rm tot}(E_\gamma)=\tau_{{\rm syn}}
(E_\gamma)+\tau_{{\rm IR}}(E_\gamma)$ is the total optical depth
for photons in the synchrotron radiation of the blob and the
infrared--microwave background, and the primed quantities are
measured in the blob frame. For the power at $\gamma$-ray
energies we adopt the value reported by the CAT experiment at
$E_\gamma = 1$ TeV (Djannati-Atai et al.\ 1998), and for the power
at X-ray synchrotron energies we take the value at
$\varepsilon=2$~keV (Pian et al.\ 1998).  For these two
energies $\eta\approx 3.6$, and from this constraint we shall now
obtain the required magnetic field as a function of the Doppler
factor for various variability time-scales.

The synchrotron spectrum at $\varepsilon'$ in the above formula 
(Eq.~\ref{eq:eta}) can be approximated analytically by
\begin{eqnarray}
\varepsilon' {{dN}\over{d\varepsilon' dt}} d\varepsilon' \approx 
{{dN}\over{d\gamma'}} d\gamma' b_{\rm syn}(\gamma'),
\end{eqnarray}
where $dN/d\gamma'$ is the electron spectrum in the blob rest frame.
$b_{\rm syn}(\gamma') = k U_B {\gamma'}^2$ is the energy loss rate of 
electrons, where $k = 4 c \sigma_T/3$, $\sigma_T$ is the
Thomson cross section, and $U_B$ is
the magnetic field energy density. 
The characteristic energy of synchrotron photons is given by
\begin{eqnarray}
\varepsilon' \approx 0.5\varepsilon_B{\gamma'}^2.
\end{eqnarray}
We use the observed low energy SED (Eq.~\ref{eq:SED}) to estimate the
blob-frame equilibrium electron spectrum
\begin{eqnarray}
{{dN}\over{d\gamma'}} \approx 
\cases{a_1 {\gamma'}^{-1.8} &
 $\gamma' \le \gamma'_{\rm b,1}$,\cr 
a_2
{\gamma'}^{-2.18} & $\gamma'_{\rm b,1} < \gamma' \le \gamma'_{\rm b,2}$,\cr
a_3
{\gamma'}^{-2.66} & $\gamma'_{\rm b,2} < \gamma' \le \gamma'_{\rm b,3}$,\cr
a_4
{\gamma'}^{-3} & $\gamma'_{\rm b,3} < \gamma' \le \gamma'_{\rm b,max}$,\cr}
\end{eqnarray}
where $\gamma'$ is the Lorentz factor in the blob frame,
\begin{eqnarray}
\gamma'_{\rm b,n} = (2\varepsilon_{\rm b,n}/D\varepsilon_B)^{1/2},
\end{eqnarray}
where $n = 1,2,3$, and $a_1={\gamma'}_{\rm b,1}^{-0.38}$,
$a_2=1$, $a_3={\gamma'}_{\rm b,2}^{0.48}$, $a_4=a_3{\gamma'}_{\rm
b,3}^{0.34}$, $\varepsilon_{\rm B} = m_{\rm e} c^2 B/B_{\rm cr}$,
$B_{\rm cr}=4.414\times 10^{13}$ G, $m_e$ is the electron rest
mass.

The synchrotron spectrum
emitted by electrons with power-law spectral index $\alpha$,
multiplied by the square of the photon energy, is given by
\begin{eqnarray}
{{dN}\over{d\varepsilon' dt'}}{\varepsilon'}^{2} \approx
{{2a k U_B{\varepsilon'}^{2}}\over{\varepsilon_B^2}} 
\left({{2\varepsilon'}\over{\varepsilon_B}}\right)^{-(\alpha+1)/2},
\label{eq:syn_spec}
\end{eqnarray}
and we use this formula for each power-law section of the electron
spectrum.

The ICS part of the Eq.~\ref{eq:eta} cannot be obtained analytically
in the general case because of the complicated form of the
Klein-Nishina cross section, and so we compute this numerically
using
\begin{eqnarray}
{{dN}\over{dE'_\gamma dt'}}{E'}_{\gamma}^{2} = 
{E'}_{\gamma}^{2} \int_{\gamma'_{\rm min}}^{\infty}  
{{dN}\over{d\gamma'}}\int_{\varepsilon'_{\rm min}}^{\infty} {{dN(\gamma', 
E'_\gamma)}\over{dt' d\varepsilon' dE'_\gamma}} d\varepsilon' d\gamma',
\label{eq:ic_spec}
\end{eqnarray} 
\noindent
where $\gamma'_{\rm min}\approx E'_\gamma/m_e c^2$,
$\varepsilon'_{\rm min} =E'_\gamma/[4\gamma' (\gamma' -
E'_\gamma/m_e c^2)]$, $E'_\gamma = E_\gamma/D$, and $dN(\gamma',
E'_\gamma)/dt' d\varepsilon' dE'_\gamma$ is the ICS spectrum (see
Eq.~2.48, in Blumenthal \& Gould~1970) produced by electrons with
Lorentz factor $\gamma'$ which scatter synchrotron photons in the
blob, and we use in this formula the soft photon spectrum 
given by Eq.~\ref{eq:photden}.

Having obtained formulae for the synchrotron power at 2~keV
(Eq.~\ref{eq:syn_spec}) and the inverse Compton power at 1~TeV
(Eq.~\ref{eq:ic_spec}), we can substitute these, together with
the optical depth reduction factors, into the equation for $\eta$
(Eq.~\ref{eq:eta}).  Setting $\eta = 3.6$, we solve this equation
numerically to obtain $B$ as a function of $D$ for various values
of $t_{\rm var}$.  The resulting allowed values of $B$ versus $D$
are plotted for $\xi=1$ as the two thick full curves (for the low
and high IRB models) in Figs.~\ref{fig:constraint}(a)--(c) for
$t_{\rm var}=2.5$~hours, 20~minutes, and 2~minutes respectively.
Fig.~\ref{fig:constraint}(d) is for $\xi=1/3$ and $t_{\rm
var}=2.5$~hours.  In the next section we further constrain the
allowed values of $B$ and $D$ by requiring the radiative time
scales be consistent with the variability time scale.


\section{Constraints from Radiative Time Scales}

The simultaneous variability observed in TeV $\gamma$-rays and
X-rays allow us to place a further constraint on the homogeneous
SSC model. The observed decrease in the observed TeV
$\gamma$-ray and X-ray fluxes may only occur if the electrons
have sufficient time to cool during the flare,
\begin{eqnarray}
{t'}_{\rm cool} \leq t_{\rm var} D.
\end{eqnarray}
This condition is required if the model is truly homogeneous,
i.e. the model we consider in the present paper which includes a
homogeneous magnetic field and constant jet direction and bulk
Lorentz factor.  In what follows, we consider the cooling time
of electrons responsible for synchrotron radiation observed at 
2~keV, i.e.
\begin{eqnarray}
\gamma' = (2 \varepsilon / D \varepsilon_B)^{1/2},
\label{eq:gamma_2kev}
\end{eqnarray}
with $\varepsilon=2$~keV.
The cooling time scale for synchrotron losses of electrons with Lorentz 
factor $\gamma'$ is given by
\begin{eqnarray}
{t'}_{\rm cool}^{\rm syn} = {{m_e c^2}\over{kU_B\gamma'}}.  
\end{eqnarray}

We next estimate the ICS cooling time.  For the soft photon spectrum we
adopt, some interactions with energetic electrons will be in the
Thomson regime while others will be in the Klein-Nishina regime with
relatively small energy loss.  To calculate the cooling time, we
neglect interactions in the Klein-Nishina regime, i.e. with
photons above $\varepsilon'_T \approx m_e c^2/\gamma'$, and use the
simple Thomson energy loss formula
\begin{eqnarray}
{t'}_{\rm cool}^{\rm ICS} \approx {{m_e c^2}\over{k
U_{\rm rad}(<\varepsilon'_T) \gamma'}},
\end{eqnarray} 
where
\begin{eqnarray}
U_{\rm rad}(<\varepsilon'_T) = \int_0^{\varepsilon'_T}
n(\varepsilon') \varepsilon' d\varepsilon'.
\end{eqnarray}
Because the optical to X--ray energy flux per log energy interval,
$F_E(E) \equiv E^2 F(E)$,
increases with energy (i.e. the photon spectrum is flatter than
$\varepsilon^{-2}$) most of the inverse Compton energy flux will
occur near
\begin{eqnarray}
E_\gamma \approx D {\gamma'}^2 \varepsilon'_T = D \gamma' m_e c^2.
\label{eq:E_2kev}
\end{eqnarray}

Thus, the ratio of the energy loss times for inverse Compton and
synchrotron loss for electrons with energies such that they
radiate synchrotron photons which are observed with energies of
$\varepsilon=2$~keV is given by the ratio, $\rho$, of emitted
energy flux per log energy interval observed at
$\varepsilon=2$~keV to the emitted energy flux per log energy
interval observed at a $\gamma$-ray energy of
\begin{eqnarray}
E_\gamma \approx (2 \varepsilon D/ \varepsilon_B)^{1/2} m_e c^2, 
\end{eqnarray}
where we have used Eqs.~\ref{eq:gamma_2kev} and \ref{eq:E_2kev}.
The ratio is then
\begin{eqnarray}
{{t'}_{\rm cool}^{\rm ICS} (\gamma') \over {t'}^{\rm syn}_{\rm
cool} (\gamma')} \approx {F_E(\varepsilon) \over {F_E(E_\gamma)
\exp[\tau_{\rm tot}(E_\gamma')]}} \equiv \rho(D,B,t_{\rm var},\xi).
\end{eqnarray}
Now, $E_\gamma=DE_\gamma'$ depends on $D$ and $B$, and $\tau_{\rm
tot}(E_\gamma')$ depends on $D$, $t_{\rm var}$, $\xi$ and $B$ and
the observed optical to X--ray SED, and so the ratio of emitted
energy flux per log energy interval depends $D$, $t_{\rm var}$,
$\xi$, $B$ and the observed SED.

The total blob-frame cooling time scale of electrons by both
processes is given by 
\begin{eqnarray}
{t'}_{\rm cool} = \left( {{1}\over{{t'}_{\rm cool}^{\rm ICS}}} + 
{{1}\over{{t'}_{\rm cool}^{\rm syn}}} \right)^{-1},
\end{eqnarray}
and this must be less than the Doppler factor multiplied by the
observed variability time,
\begin{eqnarray}
{{1}\over{{t'}_{\rm cool}^{\rm ICS}}} + 
{{1}\over{{t'}_{\rm cool}^{\rm syn}}} > {1 \over D {t}_{\rm var}}.
\label{eq:tcool_condition}
\end{eqnarray}
Using $\rho(D,B,t_{\rm var},\xi)$ we can rewrite this in two
separate ways in terms of the ICS and synchrotron cooling times
in order to investigate which process dominates the electron
cooling for specific parameters of the blob,
\begin{eqnarray}
[1+\rho(D,B,t_{\rm var},\xi)] D {t}_{\rm var} >  {{t'}_{\rm cool}^{\rm ICS}}  ,
\label{eq:t_ic_condition}
\end{eqnarray}
\begin{eqnarray}
[1+\rho(D,B,t_{\rm var},\xi)^{-1}] D {t}_{\rm var}
> {t'}_{\rm cool}^{\rm syn} .
\label{eq:t_syn_condition}
\end{eqnarray}
Each of the two equations above can be solved numerically for a
given value of $t_{\rm var}$ to obtain a constraint on $B$ as a
function of $D$, and these two constraints have been added to
Figs.~\ref{fig:constraint}(a)--(d).  In each case, a separate
curve is plotted corresponding to variability simultaneous with
the TeV $\gamma$-rays occurring at 2~keV and 0.2~keV.  In these
figures the synchrotron cooling time scale of electrons is
shorter than the observed variability time scale for parameters
above the dot-dashed curves (given by
Eq.~\ref{eq:t_syn_condition}), and the IC cooling time scale is
shorter than the variability time scale for parameters to the
left of the dotted curves (given by
Eq.~\ref{eq:t_ic_condition}). Therefore, the parameter space
allowed by the variability time scale condition
(Eq.~\ref{eq:tcool_condition}) lies above the dot-dashed curves
to the left of the dotted curves.

Before discussing the implications of these new constraints we
shall consider one further constraint.  This constraint arises
from the condition that the maximum energy of electrons
(determined by the maximum energy of synchrotron photons) must be
higher than the maximum energy of $\gamma$-ray photons in the
blob frame, i.e.
\begin{eqnarray}
\gamma'_{\rm max} m_e c^2 > E_{\rm \gamma,max} D^{-1},
\end{eqnarray}
with $\gamma'_{\rm max} = (2\varepsilon_{\rm s,max} B_{\rm cr}/m_e
c^2 B D)^{1/2}$, and the maximum energy of synchrotron photon
$\varepsilon_{\rm s,max}$.  This condition gives an upper limit
to the magnetic field in the blob
\begin{eqnarray}
B < 4.4\times 10^{-2} \varepsilon_{\rm s,max} D E_{\rm \gamma,max}^{-2}
{\rm ~G},
\end{eqnarray}
where $\varepsilon_{\rm s,max}$ is in keV and $E_{\rm \gamma,max}$ is in TeV.
The low energy part of the spectrum from Mrk 501 extends up to at
least 500~keV (Catanese et al.\ 1997) while the high energy part
extends up to at least 24~TeV (Krawczynski et al.\ 1999, Konopelko
et al.\ 1999).  The constraint obtained using these values has
been added to Figs.~\ref{fig:constraint}(a)--(d).  In the next
section we shall discuss the implications of the constraints
obtained in this and the previous sections.


\section{Discussion and Conclusion}

We have now mapped out the allowed parameter space for $\xi=1$ in
Figs.~\ref{fig:constraint}(a)--(c) for $t_{\rm var}=2.5$~hours,
20 minutes, and 2 minutes respectively;
Fig.~\ref{fig:constraint}(d) shows results for $\xi=1/3$ and
$t_{\rm var}=2.5$~hours.  Allowed combinations of $B$ and $D$ lie
on the thick solid curve corresponding to the infrared absorption
model assumed (upper curve corresponds to lower IRB model).
From the constraints on the inverse Compton and synchrotron
cooling time scales, if the 2~keV X--ray flux is observed to vary
during the flare with a similar time scale as the $\gamma$-rays, then
allowed combinations of $B$ and $D$ must be to the left of the
thick dotted curve labelled $t_{\rm i,I}$, and above the thick
dot-dashed curve labelled $t_{\rm s,I}$.  Similarly, if the
0.2~keV X--ray flux is observed to vary during the flare with a
similar time scale as the $\gamma$-rays then allowed combinations of
$B$ and $D$ must be to the left of the thin dotted curve
labelled $t_{\rm i,II}$ and above the thin dot-dashed curve
labelled $t_{\rm s,II}$.  Finally, the requirement that a $\gamma$-ray
energy be less than the energy of the electron producing it 
constrains the allowed combinations of $B$ and $D$ to be 
below the thick dashed curve.

Examining Figs.~\ref{fig:constraint}(a)--(c) we see that, as
expected, as the variability time scale decreases, the
Doppler factor and magnetic field required both increase, and the
allowed range of $\log D$ and $\log B$ decrease somewhat.
Comparing Fig.~\ref{fig:constraint}(d) for $\xi=1/3$ with
Fig.~\ref{fig:constraint}(a) for $\xi=1$, both being for $t_{\rm
var}=2.5$~hours, we see the effect of varying $\xi$
is to increase the
required magnetic field and to reduce the allowed range of $\log
D$.

Not all of the parts of the thick solid curves within the
``allowed'' parameter space will actually give a viable SSC model
as they may predict a high energy SED which may have a very
different shape from that observed.  To further delineate the
allowed values of $B$ and $D$ we calculate the spectra of 
$\gamma$--rays emerging from the blob (after photon photon absorption in
the blob synchrotron radiation) and propagating to Earth (through
the infrared background).  We have calculated the spectra for
each of the points labelled A, B, C, or D, in
Figs.~\ref{fig:constraint}(a)--(d) and normalized these to the
observed flux at 1~TeV.  The results are plotted for the lower
IRB model in Figs.~\ref{fig:specific}(a)--(d)
respectively.  Fig.~\ref{fig:specific}(e) shows the result of
using the upper IRB model (points labelled A$^\prime$,
B$^\prime$, C$^\prime$, and D$^\prime$ in Fig.~\ref{fig:constraint}a)
instead of the lower one, and we see that a
worse fit to the observed spectrum results.

We now discuss which of these spectra are consistent with the
observations.  For the longest variability time scale considered,
2.5~hours (Figs.~\ref{fig:specific}a,~\ref{fig:specific}d
and~\ref{fig:specific}e), the parameters corresponding to the
points labelled A and B are ruled out because of their poor
agreement with the observed spectrum, and point C gives the best
fit, with point D being acceptable provided further absorption
(e.g. due to photons from the accretion disk, other parts of the
jet, or a dusty molecular torus, Protheroe and Biermann 1997) can
steepen the spectrum.  As we go to shorter variability time
scales (Figs.~\ref{fig:specific}b and c) only points A are ruled
out by the data.  Table~1 summarizes the best-fitting model
parameters for each case considered.

\begin{table}
\caption{Parameters for the best fits to the Mrk 501 spectrum.}
\begin{center}
\begin{tabular}{lllcccc}
\hline
\hline
Parameter & Spectrum & $t_{\rm var}$ & $\xi$ & IRB model & $B$ (G) & $D$ \\
\hline
Fig.~3(a) C & Fig.~4(a) & 2.5 hr & 1   & lower & 0.1  & 11.5 \\
Fig.~3(b) B & Fig.~4(b) & 20 min & 1   & lower & 0.2  & 20.4 \\
Fig.~3(c) B & Fig.~4(c) & 2 min  & 1   & lower & 0.6  & 35.5 \\
Fig.~3(d) C & Fig.~4(d) & 2.5 hr & 1/3 & lower & 0.15 & 15.8 \\
Fig.~3(a) C$^\prime$ & Fig.~4(e) & 2.5 hr & 1   & upper & 0.07 & 11.2 \\
\hline
\end{tabular}
\end{center}
\label{tab1}
\end{table}

As seen in Fig.~\ref{fig:constraint}(d), which is for $t_{\rm
var}=2.5$~hours and $\xi=1/3$, the constraint from the IC cooling
time scale and the observed variability at 0.2~keV is very close
to ruling out the best-fitting spectrum (point C), and definitely
rules out the spectrum corresponding to point D.  It has been
reported by Buckley and McEnery (1999, cited by Catanese et
al. 1997 and Kataoka et al.\ 1999) that U band
observations show variability simultaneous with the TeV
$\gamma$-rays.  Such U band variability would further
constrain the allowed values of $D$ as the corresponding
constraint from the IC cooling time scale would be significantly
to the left of the curve labelled $t_{i,II}$ in
Fig.~\ref{fig:constraint}(d), ruling out the best-fitting spectrum
(point C).  Inspection of Figs.~\ref{fig:specific}(d) shows that
spectra corresponding to points A and B in
Fig.~\ref{fig:constraint}(d) are ruled out by their disagreement
with observation.  Thus, the reported U band observations would
rule out all models with $\xi \le 1/3$ implying that the allowed
blob radius is larger than $\xi c D t_{\rm var}/6$.

We would like to emphasize the importance of including
photon-photon absorption in calculations when determining the
allowed parameters of the SSC model.  For example, in previous
papers the energy at which the emitted energy flux per log energy
interval maximizes in the $\gamma$-ray region is used when
constructing constraints, as is the ratio of the peak
luminosities.  However, because of photon-photon absorption the
energy at which the emitted energy flux per log energy
interval maximizes can be significantly higher than 
the energy at which the observed energy flux per log energy
interval maximizes, and this can lead to an incorrect
determination of the allowed parameters.  For example, we show in
Fig.~\ref{fig:spectra} that the maximum in the $\gamma$-ray
spectrum shifts to lower energies by about an order of magnitude
when photon-photon absorption is included; this has
consequences for the constraint obtained by Tavecchio et
al. (1998) from a comparison of the location of the peaks in
synchrotron and IC spectra.

Finally, we point out the importance of simultaneous
multi-wavelength observations which are vital to properly
determine the physical parameters of the emission region.
To be most useful, these observations should cover the full 
energy ranges of both the low energy (synchrotron) and
high energy (Compton) parts of the SED.

\section*{Acknowledgements}
We would like to thank Anita M\"ucke for reading the manuscript
and for helpful comments.  W.B. thanks the Department of Physics
and Mathematical Physics at the University of Adelaide for
hospitality during his visit. This research is supported by a
grant from the Australian Research Council and by the Polish
Komitet Bada\'n Naukowych grant 2P03D~001~14.

\section*{References}
Aharonian F. et al., 1997, A\&A, 327, L5\\
Aharonian F. et al., 1998, A\&A, 342, 69 \\
Bednarek W., Protheroe R.J., 1997, MNRAS, 292, 646\\
Biller S. et al.,\ 1998, PRL, 80, 2992\\
Bloom S.D., Marscher A.P., 1996, ApJ 461, 657\\
Blumenthal G.R., Gould R.J., 1970, Rev. Mod. Phys., 42, 237\\
Catanese M. et al., 1997, ApJ, 487, L143\\
Catanese M. et al., 1998, ApJ, 501, 616 \\
Chadwick P.M., et al., 1998, in Proc. Veritas Workshop 
on the TeV Astrophysics of\\ \indent Extragalactic Objects, in press\\
Djannati-Atai A. et al., 1998, Abstracts of
                  the 19th Texas Symposium on Relativistic\\ \indent 
                  Astrophysics and Cosmology, held in Paris,
                  France, Dec. 14-18, 1998. Eds.:\\ \indent  J. Paul,
                  T. Montmerle, and E. Aubourg (CEA Saclay).\\
Dondi, L., Ghisellini, G., 1995, MNRAS, 273, 583\\
Fixsen D.J., Dwek E., Mather J.C., Bennett
C.L., and Shafer R.A., 1998, ApJ, 508,\\ \indent 123\\
Gaidos J.A. et al., 1996, Nature, 383, 319\\
Hauser M.G. et al., 1998, ApJ, 508, 44 \\
Hayashida N. et al., 1998, ApJ Lett., 504, L71\\
Inoue S., Takahara F., 1996, ApJ 463, 555\\
Jauch J.M., Rohrlich F., 1955, 
``The theory of photons and electrons'', Addison-Wesley\\
Kataoka J. et al., 1999, ApJ, 514, 138\\
Konopelko A. et al., 1999, Astropart.Phys. submitted,
astro-ph/990193\\
Krawczynski H. et al., 1999, in Proc BL Lac Phenomenon
Meeting, Turku, ed. L. Takalo,\\ \indent ASP Conference Series, in press\\
Krennrich F. et al., 1999, 511, 149\\
Macomb D.J. et al., 1995, ApJ, 449, L99 (Erratum
1996, ApJ 459, L111)\\
Malkan M.A., Stecker F.W., 1998, ApJ, 496, 13\\
Mastichiadis A., Kirk J.G., 1997, A\&A, 320, 19\\
Mattox, J. R. et al., 1993, ApJ, 410, 609\\
Pian E. et al., 1998, ApJ, 492, L17\\
Protheroe R.J., Biermann P.L., 1997, Astropart. Phys., 6, 293\\
Protheroe R.J., Bhat C.L., Fleury P., Lorenz E., 
Teshima M., Weekes T.C., 1998,\\ \indent  in Proc. of 25th ICRC, Durban, 
Highlight Session on Mrk 501, M.S. Potgieter\\ \indent 
 et al.\ (eds.), 9, 317\\
Punch M. et al., 1992, Nature, 358, 477\\
Quinn J. et al., 1996, ApJ, 456, L83\\
Quinn J. et al., 1999, in preparation\\
Samuelson F.W. et al., 1998, ApJ, 501, L17\\
Stanev T., and Franceschini A., 1998, ApJ, 494, L159\\
Stecker F.W., and De Jager O.C., 1998, A\&A, 334, L85\\
Svensson, R., 1987, MNRAS, 227, 403\\
Tavecchio F., Maraschi L., Ghisellini G., 1998, ApJ, 509, 608\\
Zweerink J. et al., 1997, ApJ, 490, L170\\


\newpage

{\begin{figure}[htb]
\vspace{16.5cm}
\includegraphics{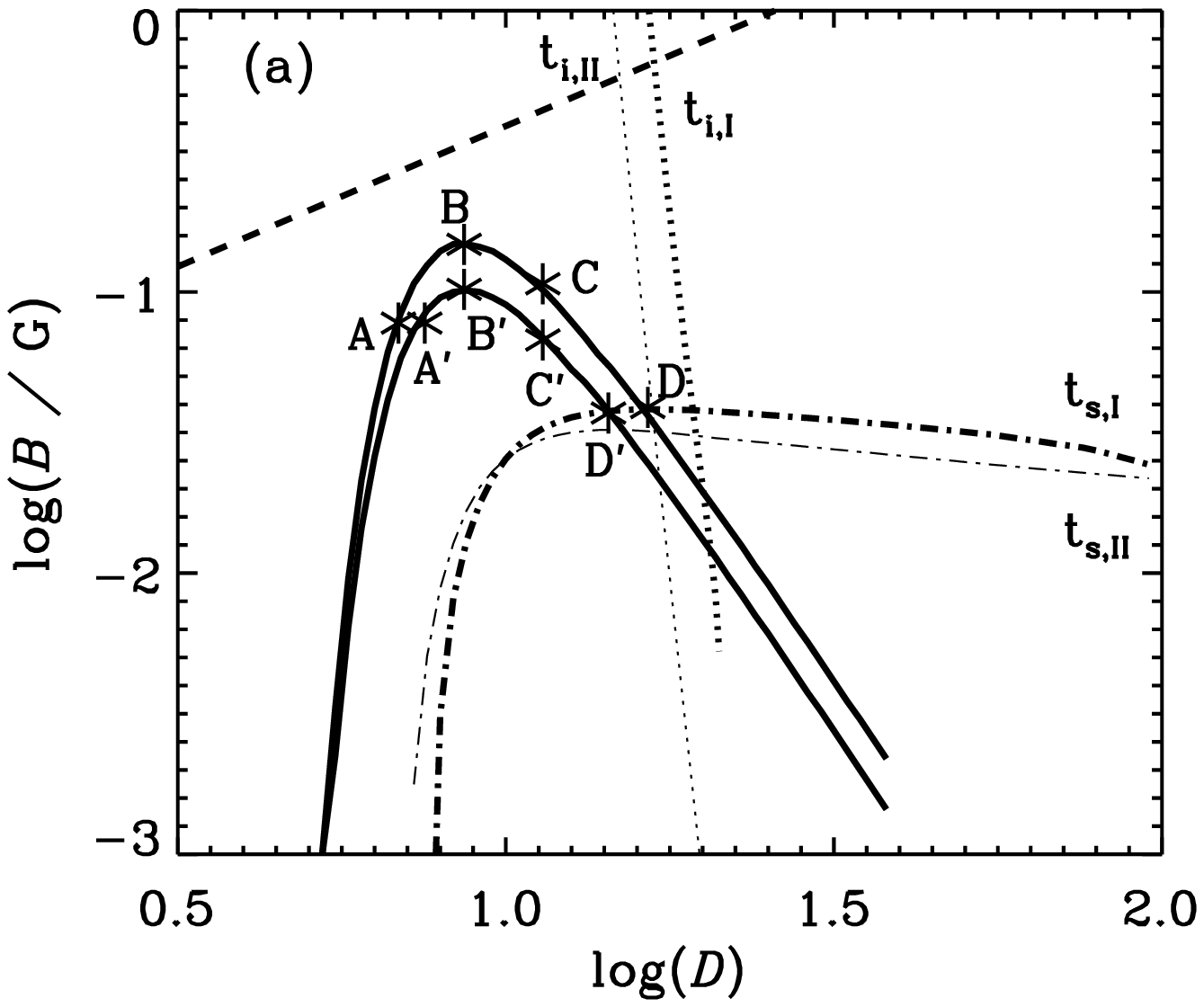}
\includegraphics{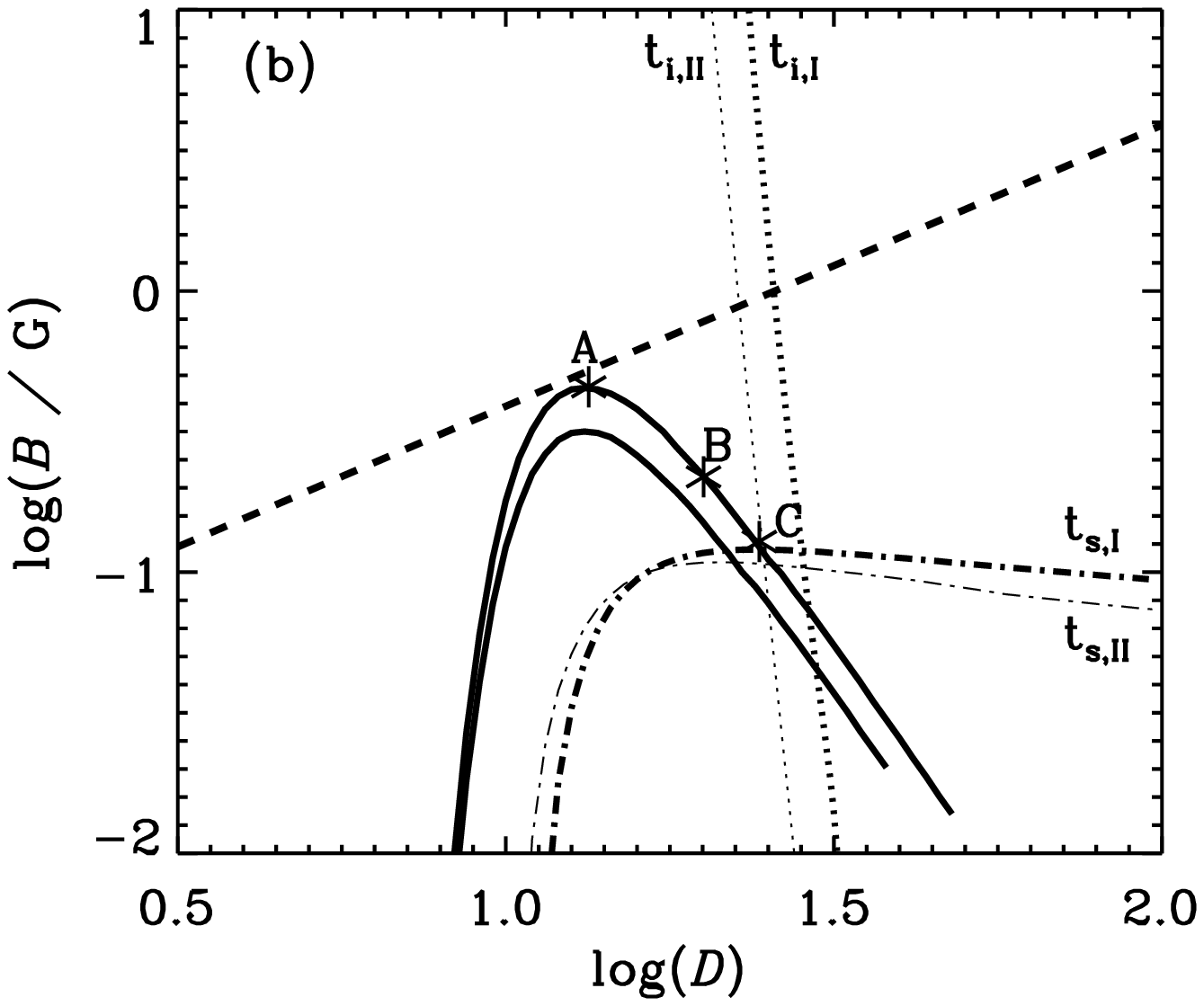}
\includegraphics{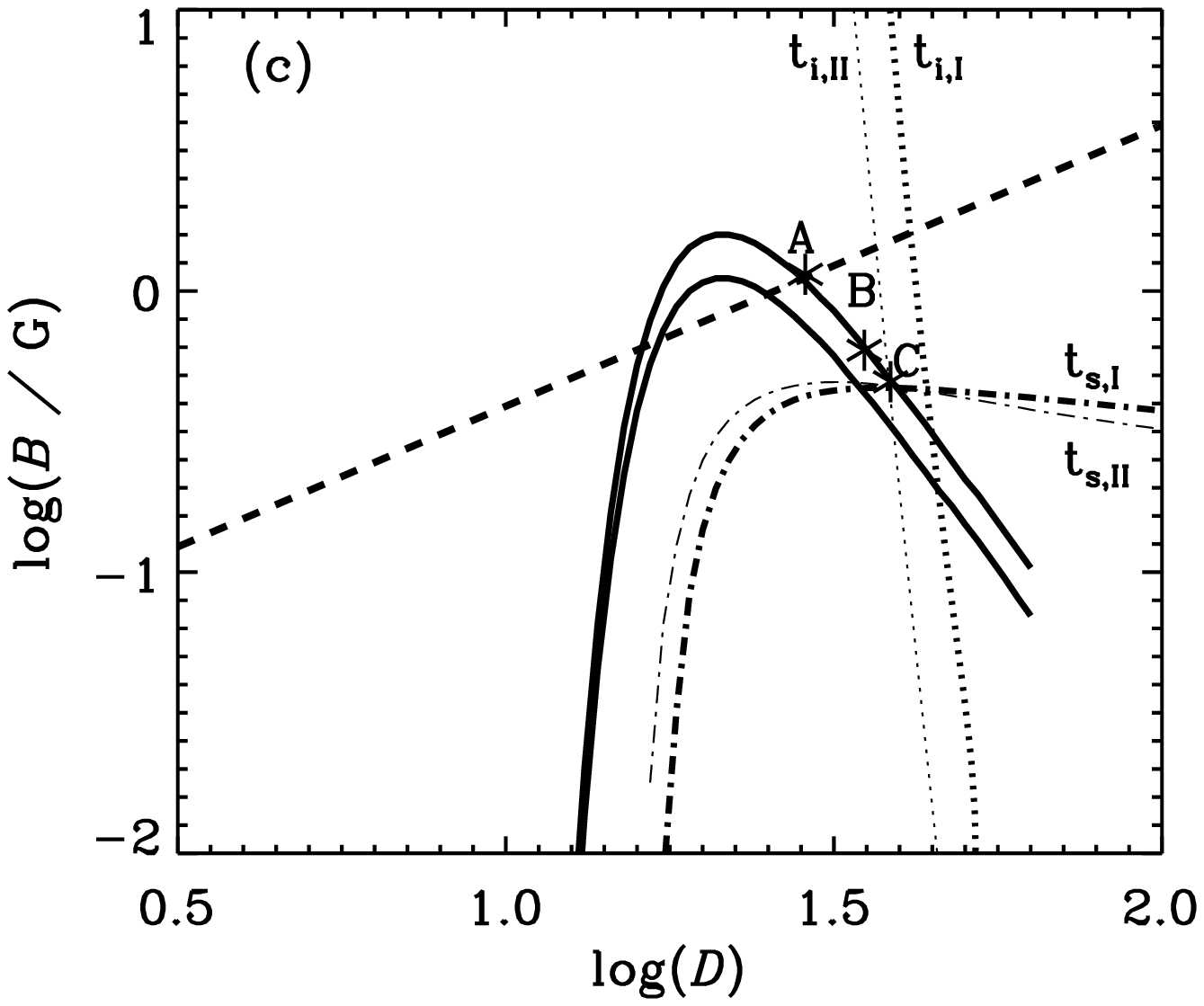}
\includegraphics{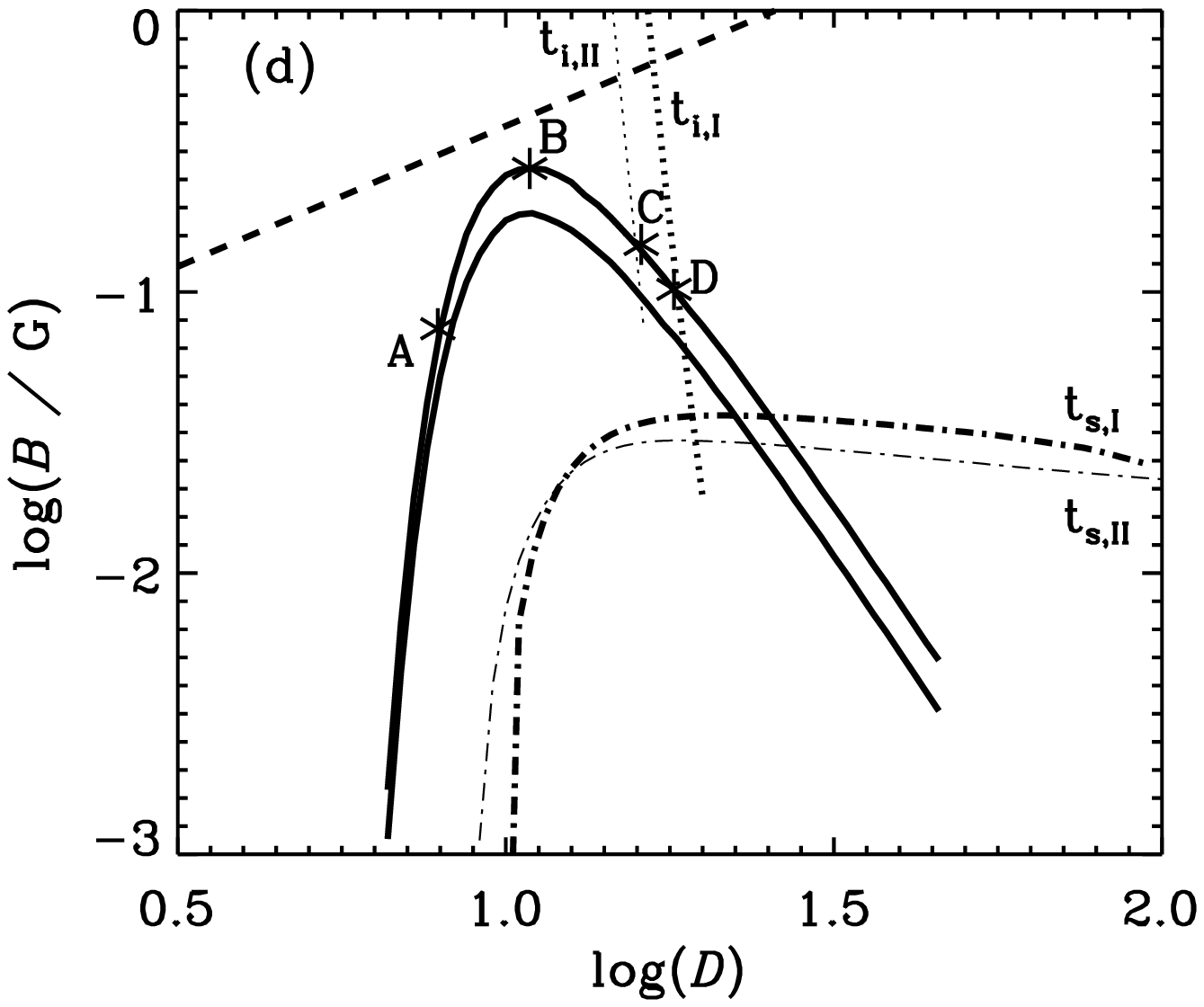}
\caption{Constraints on the parameters (magnetic field $B$ versus
Doppler factor $D$) of the blob of assumed radius $0.5 \xi
ct_{\rm var}D$ for the $\gamma$-ray flare on 15-16 April 1997 in Mrk
501 assuming different variability time scales of the flare: (a)
$t_v = 2.5$ hr and $\xi=1$, (b) $t_v = 20$ min and $\xi=1$, (c)
$t_v = 2$ min and $\xi=1$, and (d) for $t_{\rm var} = 2.5$hr and
$\xi=1/3$.  The full curves give the allowed values for $B$ and
$D$ constrained by Eq.~7 for the case of absorption of
$\gamma$-ray photons in the IRB using the lower model (upper
curve) and the upper model (lower curve), the dot-dashed curves
give the lower limit from the synchrotron cooling time scale
(Eq.~25) and the dotted lines give the upper limit from the
inverse Compton cooling time scale (Eq.~24) for electrons with
energies which produce synchrotron photons with observed energies
of 2 keV (thick lines marked by $t_{s,I}$ and $t_{i,I}$) and 0.2
keV (thin lines marked by $t_{s,II}$ and $t_{i,II}$), and the dashed
curve gives an upper limit from the maximum energies of
synchrotron and inverse Compton photons (Eq.~27). The points
labelled A, B, etc., mark the values of $B$ and $D$ for which we
have calculated $\gamma$-ray spectra and which we shall compare
with observations in Fig.~4.}
\label{fig:constraint}
\end{figure}

{\begin{figure}[htb]
\vspace{8.0cm}
\includegraphics{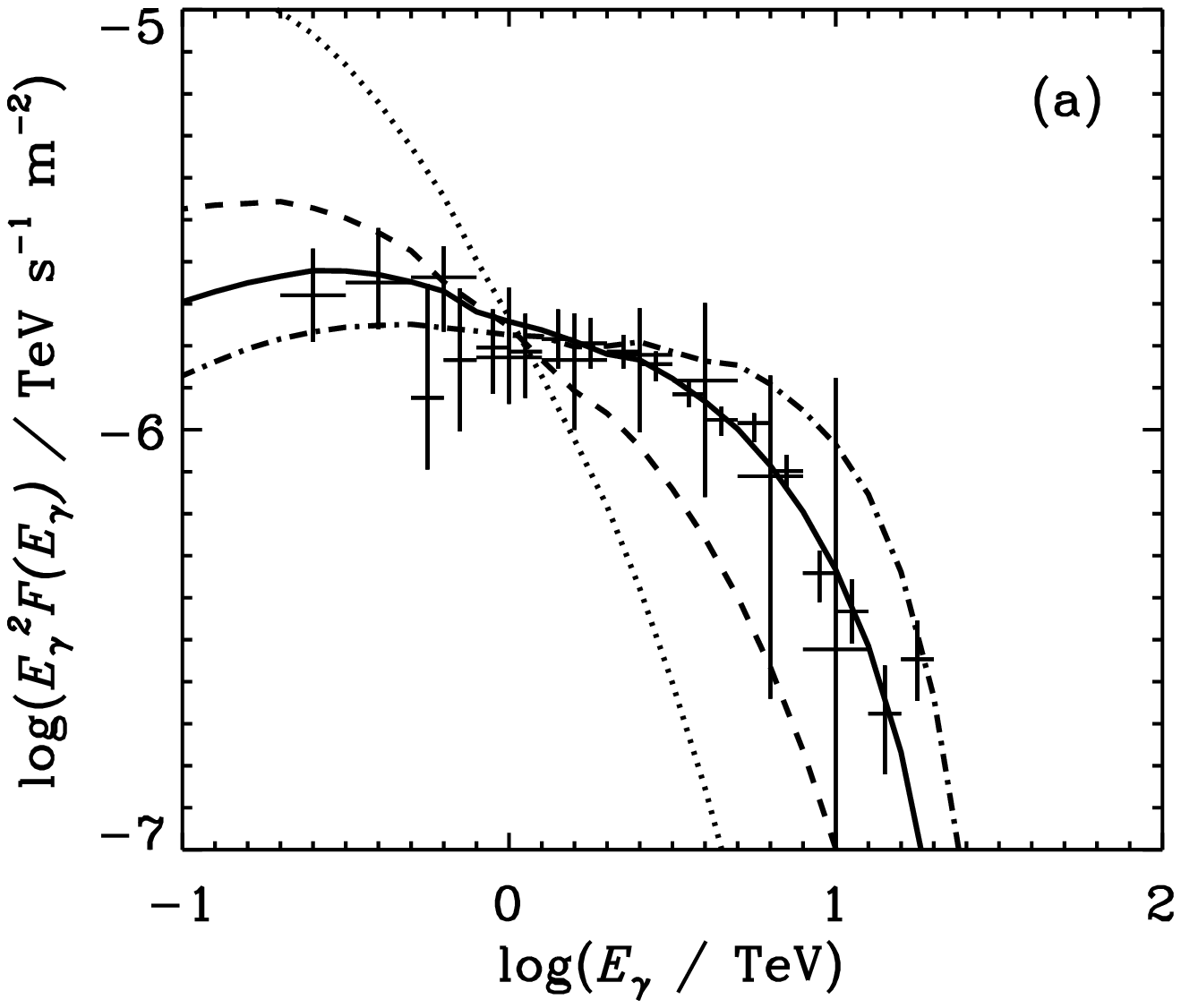}
\includegraphics{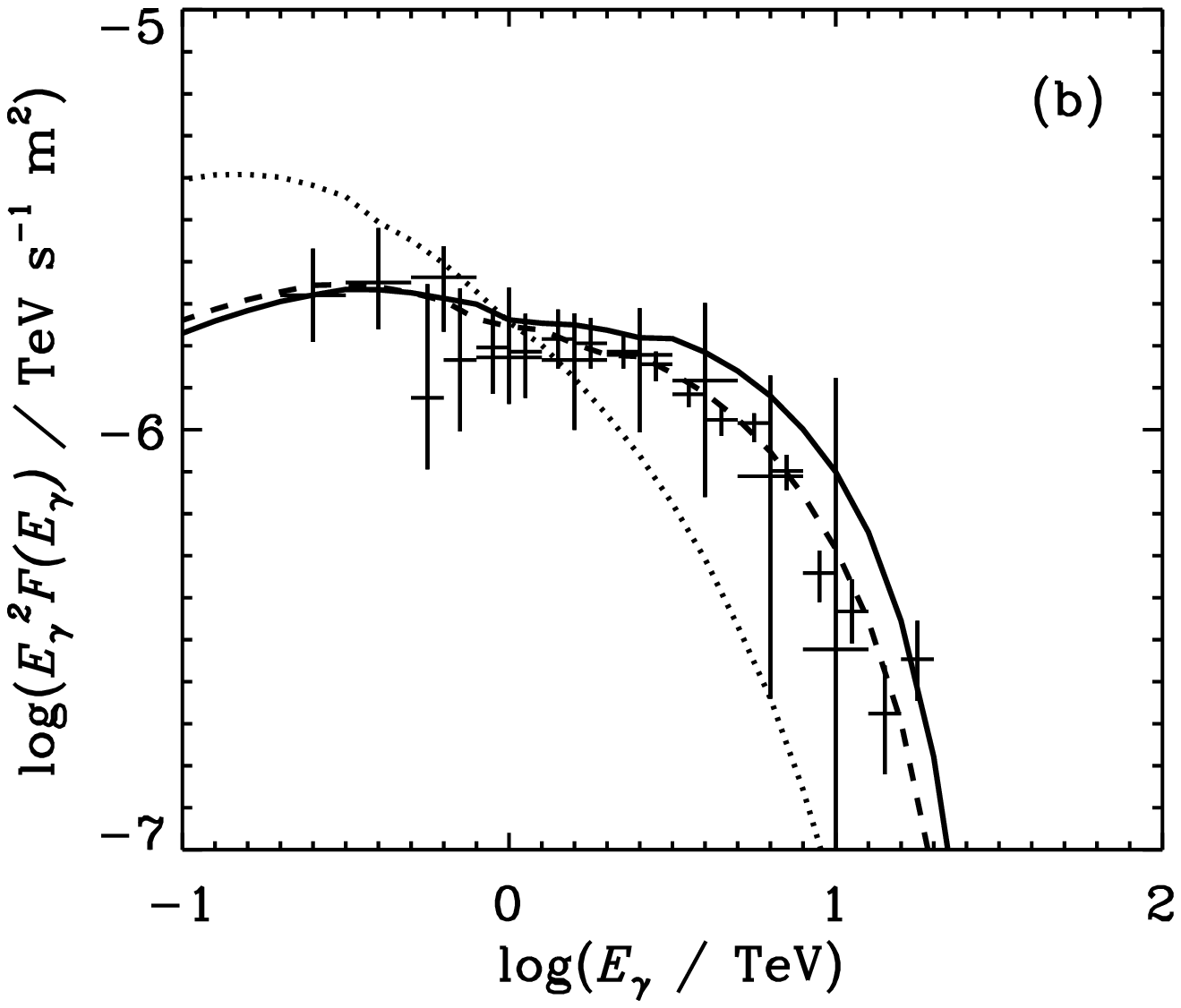}
\caption{Gamma-ray spectra computed for the specific values of
$B$ and $D$ using the lower IRB model are compared with the
observations of Mrk 501 in the 15-16 April 1997 flare by the HEGRA
(Krawczynski et al.\ 1999) and CAT (Djannati-Atai et al.\ 1998)
telescopes: (a) $t_v = 2.5$ hr and $\xi=1$, (b) $t_v = 20$ min
and $\xi=1$, (c) $t_v = 2$ min and $\xi=1$, and (d) for $t_{\rm
var} = 2.5$hr and $\xi=1/3$.  The spectra shown in (e) are
obtained using the higher IRB model, $t_v = 2.5$hr and $\xi=1$.
Specific spectra corresponding to different values of $B$ and $D$
indicated in Figs.~4(a)--(d) by A or A', B or B', C or C', and D
or D' are shown by the dotted curves, dashed curves, solid
curves, and dot-dashed curves, respectively. }
\label{fig:specific}
\end{figure}

\setcounter{figure}{3}

{\begin{figure}[htb]
\vspace{17.0cm}
\includegraphics{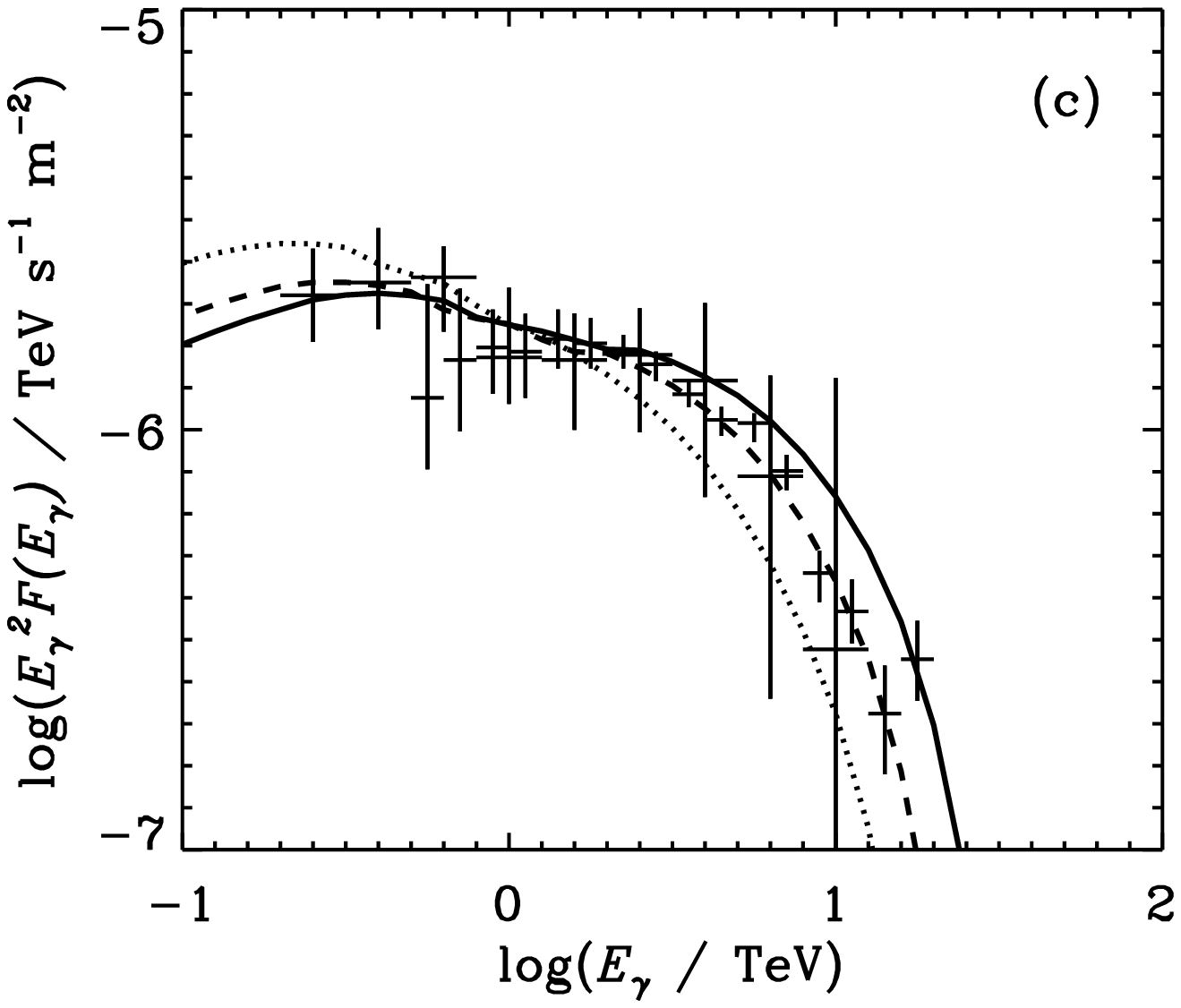}
\includegraphics{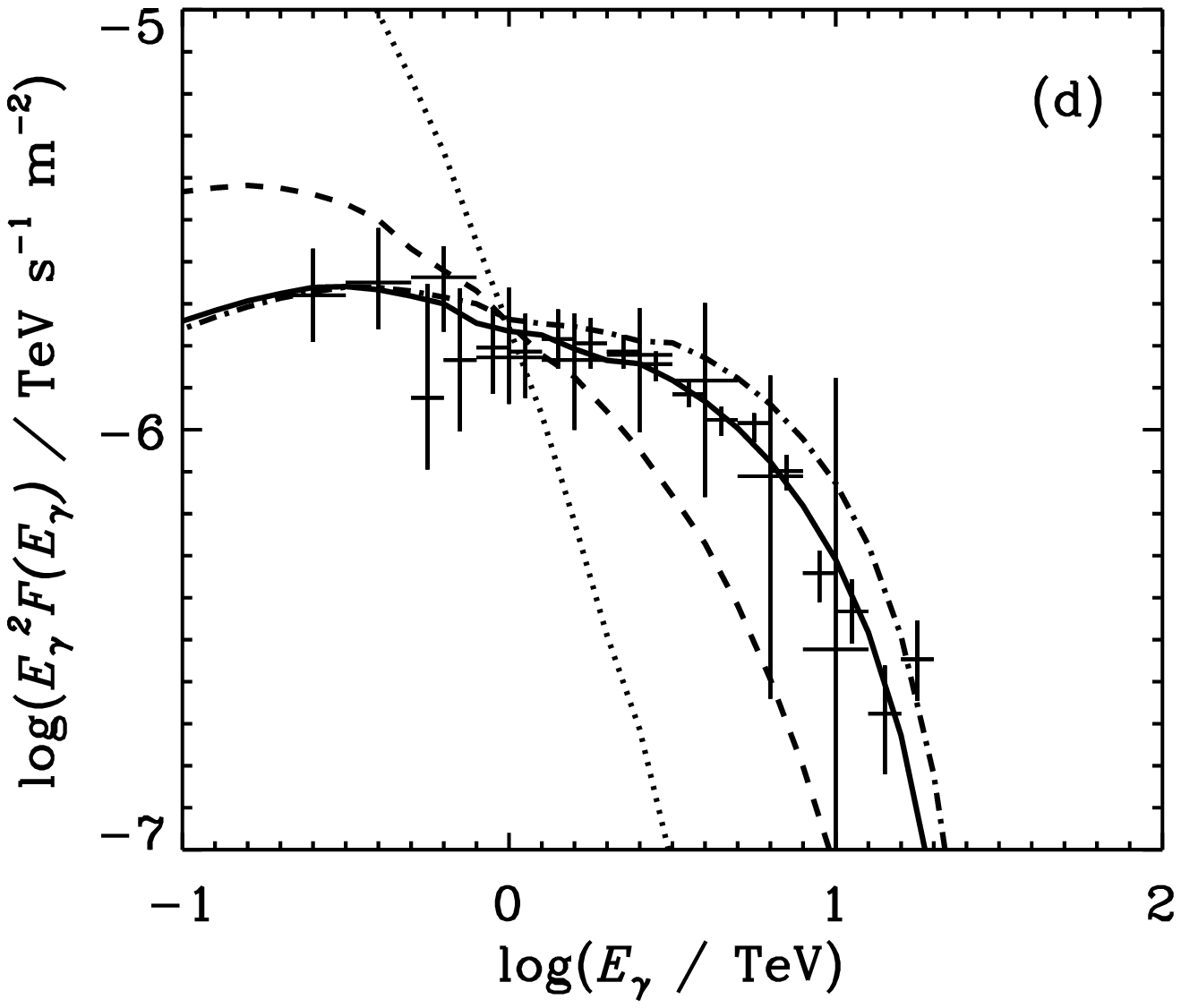}
\includegraphics{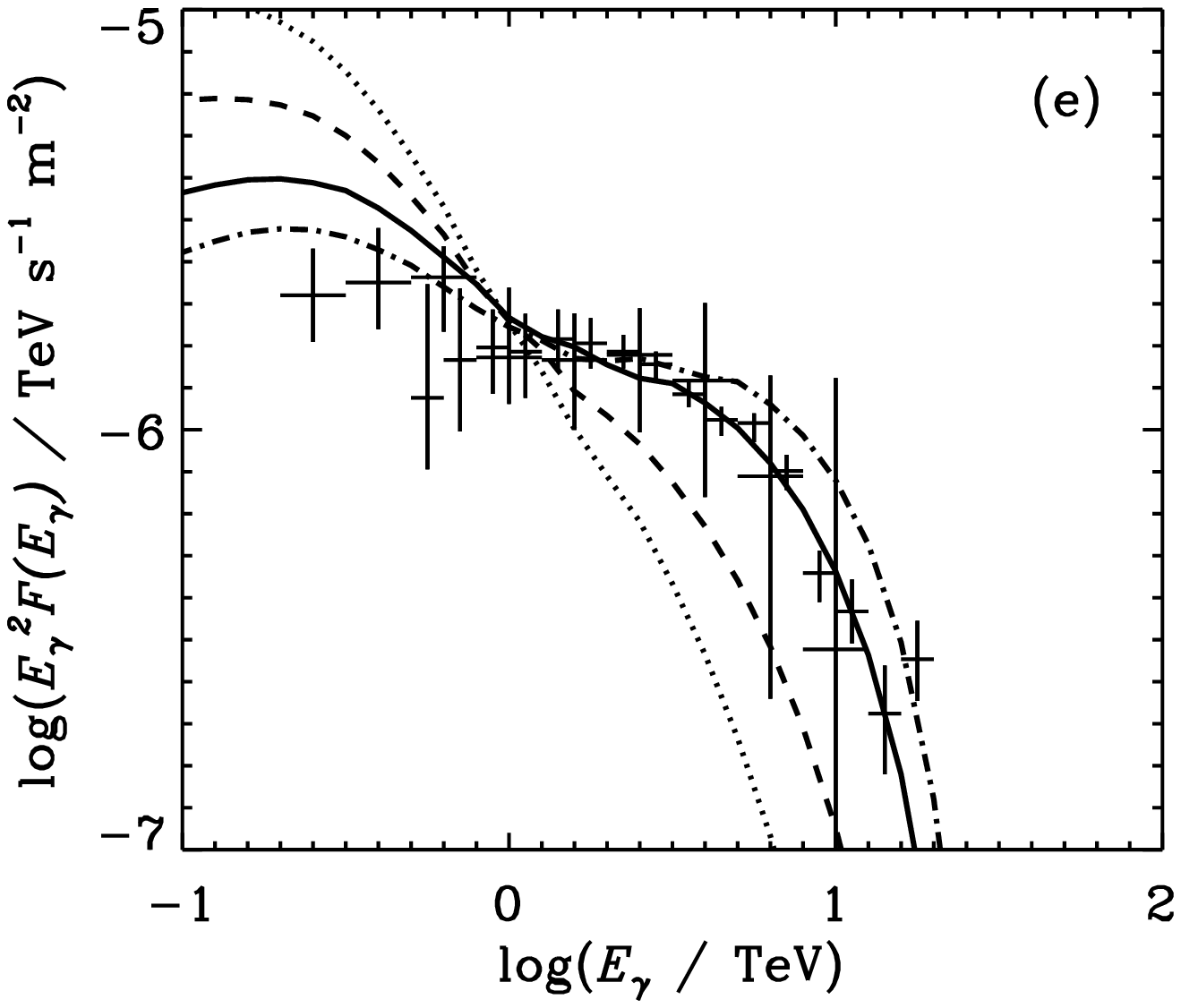}
\caption{(Continued)}
\end{figure}

{\begin{figure}[htb]
\vspace{12.0cm}
\includegraphics{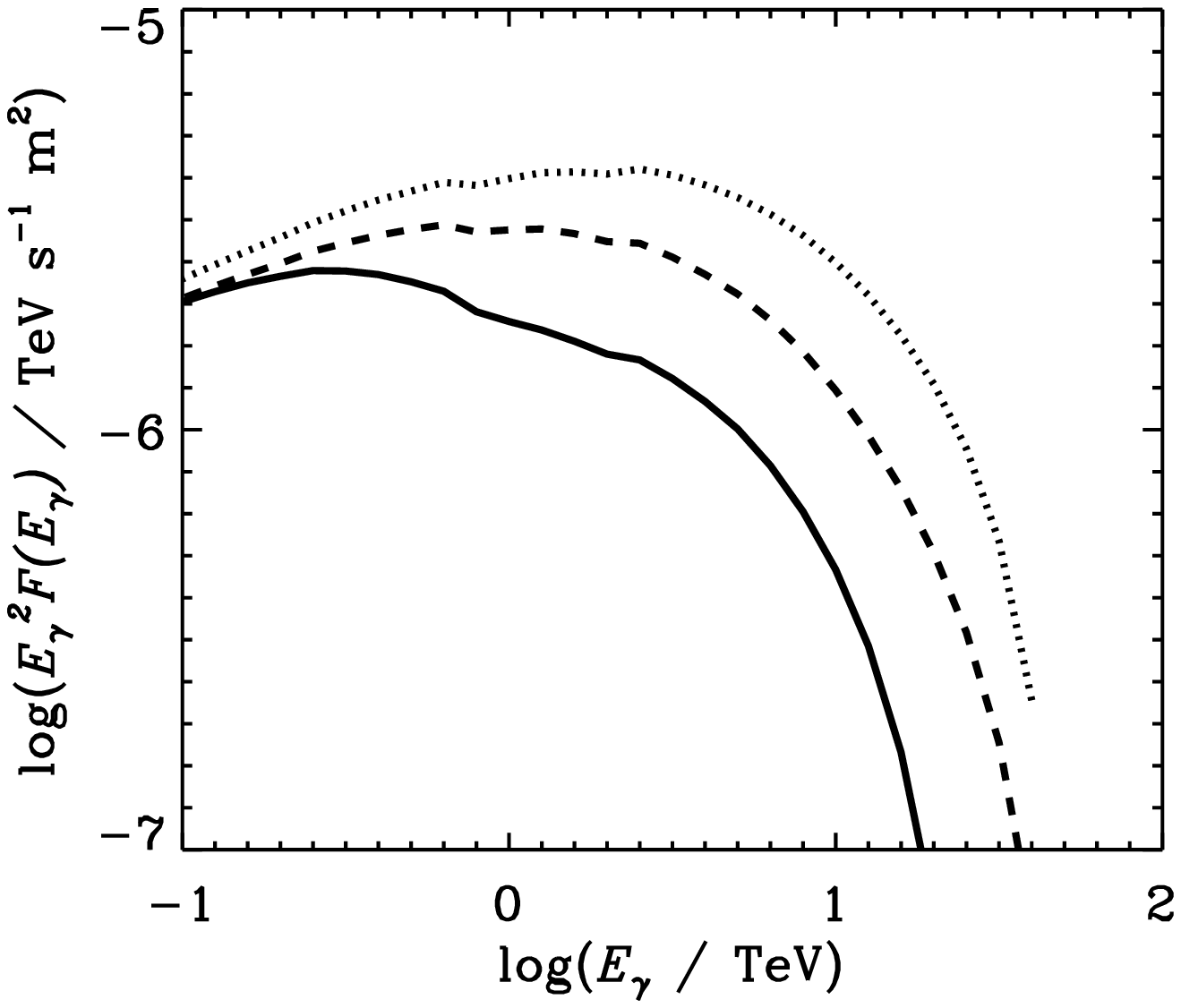}
\caption{The $\gamma$-ray spectra computed for the parameters $B$
and $D$ given by the point C in Fig.~3a. The dotted curve shows
the $\gamma$-ray spectrum produced in the blob by relativistic
electrons. The dashed curve shows the $\gamma$-ray spectrum
modified by absorption in the blob synchrotron radiation,
i.e. the spectrum emerging from the blob. The full curve shows
the $\gamma$-ray spectrum after propagation through the IRB using
the lower IRB model.}
\label{fig:spectra}
\end{figure}

\end{document}